\documentclass[a4paper,12pt]{article}
\usepackage[utf8]{inputenc}

\usepackage[left=2.5cm,right=2.5cm,bottom=4cm]{geometry}

\usepackage{mathtools,amsmath,amssymb}
\usepackage{graphicx}
\usepackage[toc,page]{appendix}

\usepackage{lmodern}
\usepackage[T1]{fontenc}
\usepackage{microtype}
\usepackage{cite}
\usepackage[margin=0.5cm]{caption}

\usepackage{hyperref}
\usepackage[punctsep]{collref}

\usepackage[usenames,dvipsnames,svgnames,table]{xcolor}
\definecolor{todocolor}{HTML}{D7E1E5}
\definecolor{lred}{HTML}{FF8888}
\usepackage[textsize=tiny, backgroundcolor=todocolor, bordercolor=todocolor, linecolor=todocolor, textwidth=3.2cm]{todonotes}
\usepackage{comment}
\usepackage{rotating}

\newcommand{\myfrac}[3][0pt]{\genfrac{}{}{}{}{\raisebox{#1}{$#2$}}{\raisebox{-#1}{$#3$}}}

\newcommand{\lax}{\mathcal{L}}

\newcommand{\inh}{v}

\newcommand{\aux}{{}}
\newcommand{\G}{G}
\newcommand{\B}{B}
\newcommand{\raps}{\Theta}

\newcommand{\K}{\mathcal{K}}

\newcommand{\ua}{\mathcal{A}}
\newcommand{\ub}{\mathcal{B}}
\newcommand{\uc}{\mathcal{C}}
\newcommand{\ud}{\mathcal{D}}
\newcommand{\udt}{{\tilde{\mathcal{D}}}}

\newcommand{\gl}{\mathfrak{gl}(2)}

\author{Rouven Frassek}
\title{Boundary Perimeter Bethe Ansatz}
\begin{document}

\begin{titlepage}
%
 
\begin{center}
\phantom{BPBA}
\vspace{0.5in}
 \textbf{\Large Boundary Perimeter Bethe Ansatz}
\\
\vspace{1in}
{\large Rouven Frassek}
\\[0.2in]
Institut des Hautes \'{E}tudes Scientifique, France
 \vspace{.8in}
    \begin{center}
  \includegraphics[width=0.4\linewidth]{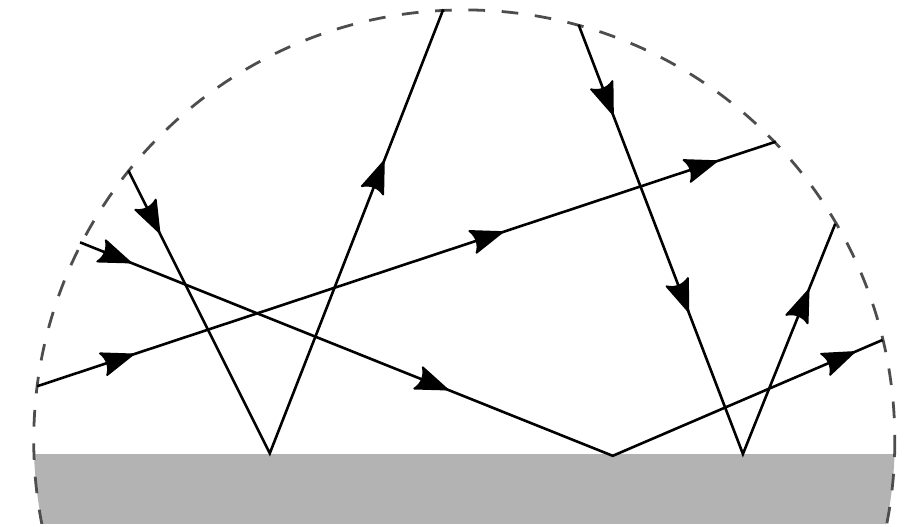}
\end{center}
 \end{center}
 \vspace{.4in}
\begin{center}
\textbf{\large Abstract}
\end{center}
\begin{center}
\begin{minipage}{400pt}
 \noindent  
We study the partition function of the six-vertex model in the rational limit on arbitrary Baxter lattices with reflecting boundary. Every such lattice is interpreted as an invariant of the twisted Yangian. This identification allows us to relate the partition function of the vertex model to the Bethe wave function of an open spin chain. We obtain the partition function in terms of creation operators on a reference state from the algebraic Bethe ansatz and as a sum of permutations and reflections from the coordinate Bethe ansatz.
\end{minipage}
\end{center}
\end{titlepage}

\vfill
\newpage
\setcounter{tocdepth}{1}
\tableofcontents
\vspace{2cm}

\section{Introduction}\label{sec:intro}

In 1987, R.~J.~Baxter suggested an intriguing relation between the partition function of the six-vertex model on an arbitrary (Baxter) lattice and the Bethe wave function of the {\small XXZ} spin chain \cite{Baxter1987}. More precisely, for a lattice of $N$ lines crossing in an arbitrary way the partition function was identified up to a simple factor with the coordinate wave function of a half-filled inhomogeneous {\small XXZ} spin chain of length $L=2N$, cf.~\cite{Bethe1931}. Depending on the lattice shape Baxter gave a prescription to identify the $2N$ inhomogeneities and $N$ Bethe roots  of the spin chain with the $N$ rapidities of the lattice model. Furthermore, he noticed that this identification solves the Bethe equations of the inhomogeneous {\small XXZ} spin chain exactly. The final expression for the partition function  emerging in this way is given explicitly by a sum of $N!$ terms. The method described is referred to as the perimeter Bethe ansatz.

Baxter's perimeter Bethe ansatz was reformulated in the framework of the quantum inverse scattering method \cite{Frassek2014}. In particular, for the rational limit it was shown that the eigenstates of the spin chain, whose entries correspond to partition functions of a given Baxter lattice with different external state configurations, are Yangian invariant. This allowed to generalise Baxter's perimeter Bethe ansatz to symmetry algebras of higher rank and different representations; we refer the reader to \cite{Faddeev1996} for an introduction to the quantum inverse scattering method. Again, as in \cite{Baxter1987}, the inhomogeneities and Bethe roots have to be identified with the rapidities of the lattice model. Moreover, it was pointed out in \cite{Frassek2014} that these identifications yield solutions to certain Bethe type equation which imply the ordinary Bethe equations.

A natural question that arises is whether the perimeter Bethe ansatz can be generalised to integrable models with boundaries, or formulated differently for the rational limit: What is the lattice model that corresponds to the invariants of a twisted (boundary) Yangian? 
Integrable models with boundary are well studied and enjoy increasing popularity. The open {\small XXZ} spin chain with diagonal boundary conditions was solved in~\cite{Alcaraz1987,Gaudin2014} using the coordinate Bethe ansatz. Shortly after, E.~K.~Sklyanin formulated the quantum inverse scattering method for boundary models and solved the open {\small XXZ} chain using the algebraic Bethe ansatz~\cite{Sklyanin:1988yz}. Finally, also the quantum groups underlying the boundary Yang-Baxter equation \cite{Cherednik1984a} are well studied. For the twisted Yangian appearing in the rational limit see~\cite{Olshanskii1992,Molev:2007} but also~\cite{MacKay2005} for an excellent overview. 

In the following we combine these methods for integrable spin chains with boundaries and generalise the perimeter Bethe ansatz.  We find that the invariants of the twisted Yangian are related to the partition functions of Baxter lattices with boundary as shown in Figure~\ref{fig:lattice}. After deriving the Bethe equations for the invariants of the twisted Yangian, we obtain the partition function in terms of creation operators on a reference state in the framework of the algebraic Bethe ansatz. Furthermore, we derive the inhomogeneous wave function for the open spin chain as familiar from the coordinate Bethe ansatz and relate it to the partition function. 

The paper is organised as follows: Instead of starting with the study of twisted Yangian invariants in the framework of the algebraic Bethe ansatz we first introduce the lattice model and define its partition function. In Section~\ref{sec:twistedY}, we show that the defined lattices yield invariants of the twisted Yangian employing its so-called {\small RTT}-realisation. We introduce elementary invariants and discuss how their tensor product and multiplication by R-matrices yield further invariants. In Section~\ref{sec:ABA}, we formulate the algebraic Bethe ansatz for twisted Yangian invariants and express the partition function in terms of creation operators on a reference state. Finally, in Section~\ref{sec:CBA}, we employ the coordinate  representation of the Bethe vectors (coordinate Bethe ansatz) to rewrite the partition function as a sum of $2^N\times N!$ terms and end with a conclusion.  

\section{Baxter lattice with boundary}\label{sec:BBL}
\begin{figure}[ht!]
\begin{center}
  \includegraphics[width=0.6\linewidth]{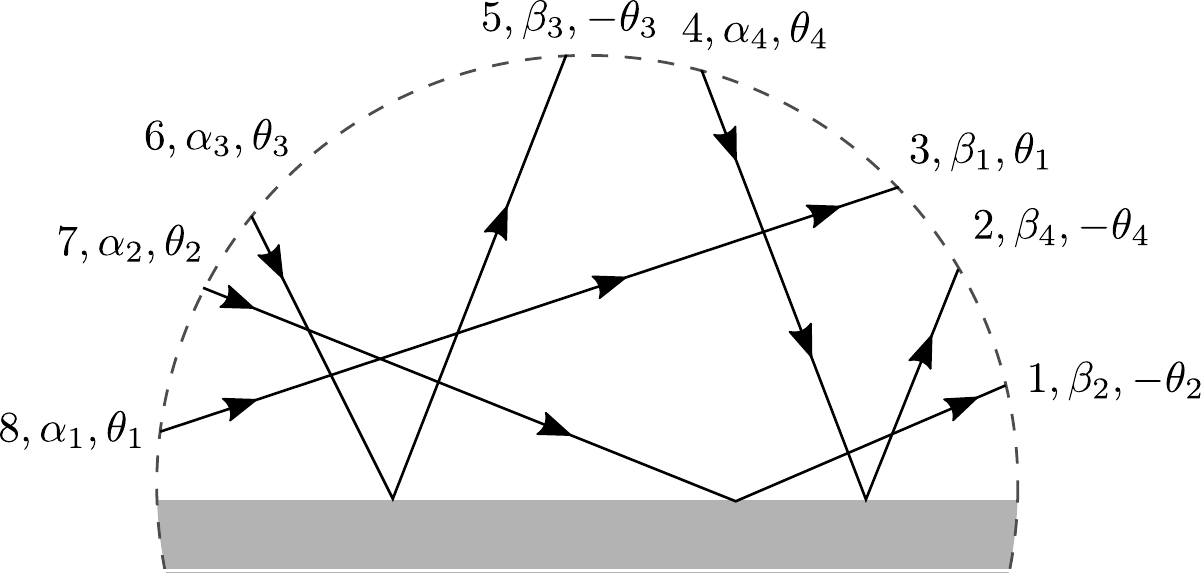}
\end{center}
  \caption{Baxter lattice with boundary for $N=4$ lines and the sets $\G=((8,3),(7,1),(6,5),(4,2))$ and $\B=\{2,3,4\}$. The indices $\alpha_i$ and $\beta_i$ label the states at the edges on the perimeter associated to a rapidity $\pm\theta_i$ depending on the set $B$.}
  \label{fig:lattice}
\end{figure}

We consider a lattice model defined on the domain of a half-disk consisting of $N$ lines that are enclosed by the arc segment (perimeter). Lines get reflected on the diameter segment (boundary). Three lines should not meet in one point and none of the points where lines get reflected on the boundary should coincide. We label the points where the lines meet the arc segment counterclockwise from $1,\ldots,2N$. In this way we assign two integers to each line $(i,j)$ with $2N\geq i>j\geq 1$ and define the ordered set
\begin{equation}\label{eq:G}
\G=((i_1,j_1),\ldots,(i_N,j_N))\,,
\end{equation} 
where $i_1>i_2>\ldots>i_N$. Here we implicitly associated an integer $k=1,\ldots,N$ to every line $(i_k,j_k)$, cf.~\cite{Frassek2014}. To distinguish the lines that get reflected from the diameter segment from the ones that do not, we introduce the set
\begin{equation}\label{eq:B}
B\subseteq\{1,2,\ldots,N\}\,.
\end{equation} 
It contains the lines $k\in\B$ that get reflected at the diameter segment. The complement $\bar\B=\{1,2,\ldots,N\}/\B$ contains the lines that do not get reflected. All lines $(i_k,j_k)$ are oriented,  which is indicated by an arrow pointing towards the end point $j_k$. Furthermore, we assign a rapidity $\theta_k$ to every oriented line $k$ and introduce the set of rapidities 
\begin{equation}
\raps=\{\theta_1,\ldots,\theta_N\}\,.
\end{equation} 
It is important to note that the rapidities associated to lines $k\in\B$ change sign $\theta_k\rightarrow -\theta_k$ when reflected at the boundary, cf. \eqref{eq:krefl}. An example of a lattice described here can be found in Figure~\ref{fig:lattice}.

Next we assign Boltzmann weights to the lattice. To do so we first associate state labels  $\alpha,\beta,\gamma,\delta=1,2$ to the edges of the lattice. 
The weights of the four-valent vertices are determined by the entries of an R-matrix
\begin{equation}
 \langle\alpha_1\alpha_2|R(\theta_1-\theta_2)|\beta_1\beta_2\rangle=
  \begin{aligned}
  \includegraphics[scale=0.80]{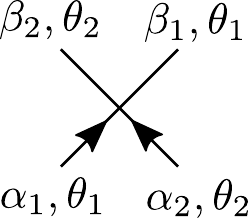}\,.
 \end{aligned}
\end{equation} 
Here we use the bra-ket notation to denote the basis vectors $|1\rangle
= \bigl(\begin{smallmatrix}1\\0\end{smallmatrix}\bigr)$ and $|2\rangle
= \bigl(\begin{smallmatrix}0\\1\end{smallmatrix}\bigr)$ while $|\beta_1\beta_2\rangle=|\beta_1\rangle\otimes|\beta_2\rangle$ and $\langle\alpha_1\alpha_2|=\langle\alpha_1|\otimes\langle\alpha_2|$ denote four-vectors built from the tensor product. The R-matrix is a $4\times 4$-matrix with $R(\theta):\mathbb{C}^2\otimes\mathbb{C}^2\rightarrow\mathbb{C}^2\otimes\mathbb{C}^2$ and has to satisfy the Yang-Baxter equation
\begin{equation}\label{eq:ybe}
R_{12}(\theta_1-\theta_2)R_{13}(\theta_1-\theta_3)R_{23}(\theta_2-\theta_3)=R_{23}(\theta_2-\theta_3)R_{13}(\theta_1-\theta_3)R_{12}(\theta_1-\theta_2)\,.
\end{equation} 
Here $R_{12}$ acts trivially in space $3$, i.e. $R_{12}(\theta)=R(\theta)\otimes\mathbb{I}$ with the $2\times 2$ identity matrix $\mathbb{I}$, and similarly for $R_{13}$ and $R_{23}$. 
We concentrate on the six-vertex model in the rational limit such that at each vertex the weights are determined by the R-matrix
\begin{equation}\label{eq:rmatrix}
 R(\theta)=\frac{1}{\theta+1}\left(\begin{array}{cccc}
             \theta+1&0&0&0\\
             0&\theta&1&0\\
             0&1&\theta&0\\
             0&0&0&\theta+1\\           
            \end{array}
\right)\,.
\end{equation} 
In addition to the four-valent vertices in the bulk we also associate Boltzmann weights to the vertices where a line is reflected from the boundary. 
The Boltzmann weights of the boundary vertices are determined by a K-matrix
 \begin{equation}\label{eq:krefl}
 \langle\alpha|K(\theta)|\beta\rangle= 
 \begin{aligned}
  \includegraphics[scale=0.80]{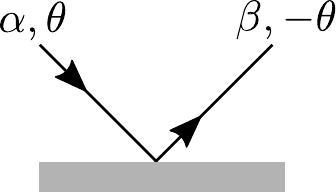}\,.
 \end{aligned}
\end{equation} 
The K-matrix is a $2\times 2$-matrix and has to satisfy the boundary Yang-Baxter equation
\begin{equation}\label{eq:bybe}
 R(\theta_1-\theta_2)K_1(\theta_1)R(\theta_1+\theta_2) K_2(\theta_2)=  K_2(\theta_2)R(\theta_1+\theta_2)K_1(\theta_1)R(\theta_1-\theta_2)\,.
\end{equation} 
Here we use the same notation as for the R-matrix such that the K-matrix\, $K_1(\theta)=K(\theta)\otimes\mathbb{I}$ acts trivially in the second space. As indicated in \eqref{eq:krefl} and remarked earlier, the sign of the rapidity changes when a line is reflected at the boundary. We restrict ourselves to the case of the diagonal K-matrix 
\begin{equation}\label{eq:kmatrix}
 K(\theta)=\frac{1}{q+\theta}\left(\begin{array}{cc}
                    q+\theta&0\\
                    0&q-\theta
                   \end{array}
\right)\,,
\end{equation} 
which is a solution to the boundary Yang-Baxter equation \eqref{eq:bybe}.
Finally, we also introduce Boltzmann weights for single lines that neither get reflected at the boundary of the lattice nor cross any other lines. These weights do not depend on the rapidities and are defined as 
\begin{equation}\label{eq:idmatrix}
 \delta_{\alpha\beta}=
 \langle\alpha|\beta\rangle= 
 \begin{aligned}
  \includegraphics[scale=0.80]{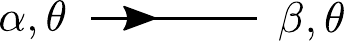}\,.
 \end{aligned}
\end{equation} 

We are now in the position to define the partition function of a Baxter lattice with boundary for a given lattice denoted by the sets $\G$ and $\B$ and a fixed configuration of states at the edges on the perimeter labelled by 
\begin{equation}
 \boldsymbol\alpha=\{\alpha_1,\ldots,\alpha_N\}\,,\qquad 
 \boldsymbol\beta=\{\beta_1,\ldots,\beta_N\}\,.
\end{equation} 
Here a pair $(\alpha_i,\beta_i)$ labels the state at the in- and outpointing arrows on the edges of the $i$-th line, cf.~Figure~\ref{fig:lattice}.
The partition function is defined as the product of internal vertices with the subsequent sum over all internal state configurations 
\begin{equation}\label{eq:defZ}
 Z(\G,\B,\raps,\boldsymbol\alpha,\boldsymbol\beta)=\sum_{\substack{\text{internal}\\ \text{state}\\ \text{config.}}}\,\prod_{\text{vertices}} (\text{Boltzmann weight})\,.
\end{equation} 
As for the six-vertex model on an ordinary Baxter lattice the ice-rule applies, i.e. the partition function vanishes if the total number of external states on the arc of type ``$1$'' (``$2$'') at edges with ingoing arrows $i$ is not equal to the total number of states ``$2$'' (``$1$'') at edges with outgoing arrows~$j$, cf.~\cite{Frassek2014}. Further we note that our choice of normalisation of the R-matrix \eqref{eq:rmatrix} and the K-matrix \eqref{eq:kmatrix} ensures that 
\begin{equation}\label{eq:defZ0}
 Z(\G,\B,\raps,\boldsymbol\alpha_0,\boldsymbol\beta_0)=1\,,
\end{equation} 
where $ \boldsymbol\alpha_0=\{1,\ldots,1\}$ and $\boldsymbol\beta_0=\{1,\ldots,1\}$.

Finally, we note that the Yang-Baxter equation \eqref{eq:ybe} together with the boundary Yang-Baxter equation \eqref{eq:bybe} ensure that the partition function \eqref{eq:defZ} remains unchanged when moving a line through a four-valent vertex or a boundary point through another one without changing the order of the end points. In particular, the Yang-Baxter equation in \eqref{eq:ybe} also holds when $\theta_i\rightarrow-\theta_i$ for any $i=1,2,3$ and thus for any combination of reflected and non-reflected lines. The invariance of the partition function under the Yang-Baxter equation is usually referred to as $Z$-invariance \cite{Baxter1987}, see also \cite{Zamolodchikov1990}. Here, in the case of the described boundary model, we additionally have to require that the boundary Yang-Baxter equation is satisfied at the diameter segment.

\section{Invariants of the twisted Yangian}\label{sec:twistedY}
In the following we discuss how the partition function of a given boundary Baxter lattice with different external state configurations relates to invariants of a twisted Yangian. 

Let us first introduce the double-row monodromy and a set of eigenvalue equations which serve as a definition of the invariants of the twisted Yangian relevant to describe the partition function of Baxter lattices with boundary. As usual when studying {\small XXX} spin chains in the framework of the quantum inverse scattering method we introduce the $\gl$-invariant Lax operator acting on the tensor product of two fundamental representations \cite{Faddeev1996}. It reads
\begin{equation}\label{eq:lax}
 \mathcal{L}_{i}(z)=z+\sum_{a,b=1}^2e^{(i)}_{ab}\otimes e_{ba}\,,
\end{equation} 
where $e_{ab}$ are the elementary $2\times 2$ matrices with $\left(e_{ab}\right)_{cd}=\delta_{ac}\delta_{bd}$. The Lax operator $ \mathcal{L}_{i}$ acts non-trivially on the $i$-th site of the spin chain and in the auxiliary space. It coincides with the R-matrix introduced in \eqref{eq:rmatrix} up to a normalisation and  satisfies the unitarity relation
\begin{equation}\label{eq:unitarity}
\lax_i(z)\lax_i(-z)=1-z^2\,.
\end{equation} 
The Lax operator $ \mathcal{L}_{i}$ in \eqref{eq:lax} and the K-matrix
\begin{equation}
 \K(z)=(q+z)K(z)\,,
\end{equation} 
cf.~\eqref{eq:kmatrix}, are the basic building blocks of the  double-row monodromy of the open {\small XXX} Heisenberg spin chain \cite{Sklyanin:1988yz}. It can be defined as
\begin{equation}\label{eq:doublerow}
 \mathcal{U}(z)=\mathcal{M}(z)\K(z)\hat{\mathcal{M}}(z)\,.
\end{equation} 
The double-row monodromy $\mathcal{U}$ is a $2^L\times 2^L$-matrix in the quantum space and a $2\times 2$-matrix in the auxiliary space.
While the boundary K-matrix only acts non-trivially in the auxiliary space, the single-row monodromies $\mathcal{M}$ and $\hat{\mathcal{M}}$ also act non-trivially on the quantum space. They are built from the tensor product of the Lax operators at sites $i$ for $i=1,\ldots,L$ and multiplication in the auxiliary space as
\begin{equation}
 \mathcal{M}(z)=\lax_{1}(z-\inh_1)\cdots \lax_{L}(z-\inh_L)\,,\qquad\text{and}\qquad \hat{\mathcal{M}}(z)=\lax_{L}(z+\inh_L)\cdots \lax_{1}(z+\inh_1)\,.
\end{equation} 
As a consequence of the Yang-Baxter equation \eqref{eq:ybe} and the boundary Yang-Baxter equation \eqref{eq:bybe}, the double-row monodromy \eqref{eq:doublerow} satisfies the boundary Yang-Baxter equation
\begin{equation}\label{eq:ubybe}
 R(x-y)\left(\mathcal{U}(x)\otimes \mathbb{I}\right)R(x+y)\left(\mathbb{I}\otimes \mathcal{U}(y)\right)= \left(\mathbb{I}\otimes \mathcal{U}(y)\right)R(x+y)\left(\mathcal{U}(x)\otimes \mathbb{I}\right)R(x-y)\,.
\end{equation} 
The boundary Yang-Baxter equation provides the definition of the twisted Yangian \cite{Olshanskii1992,Molev:2007}.
In order to establish the relation between the open Heisenberg chain and the boundary Baxter lattice introduced in Section~\ref{sec:BBL} we recall that a given lattice is labelled by the ordered set  $\G$  of start and end points of the lines and the set $\B$ labelling the lines that get reflected from the diameter segment, cf. \eqref{eq:G} and \eqref{eq:B}. As we will see, it is convenient to introduce the Lax operator with the conjugate representation $e_{ab}\rightarrow -e_{ba}$ at site $i$ of the quantum space, i.e. 
\begin{equation}
  \bar{\mathcal{L}}_{i}(z)=z+1-\sum_{a,b=1}^2e^{(i)}_{ab}\otimes e_{ab}\,.
\end{equation} 
For $\gl$ these representations are equivalent  and the Lax operators are related through a similarity transformation at site $i$ via
\begin{equation}\label{eq:strans}
 \bar{\mathcal{L}}_{i}(z)=S_i\,{\mathcal{L}}_{i}(z)\,S_i^{-1}\,,\qquad\text{with}\qquad S_i=\left(\begin{array}{cc} 
 0&1\\
 -1&0
\end{array}\right)_i\,.
\end{equation} 
Thus, the Lax operator $\bar{\mathcal{L}}_{i}$ satisfies the same unitarity relation as the Lax operator $\lax_i$, cf.~\eqref{eq:unitarity}. 

We are interested in double-row monodromies that contain the Lax operators $\lax$ as well as $\bar{\lax}$. More precisely, the double-row monodromy relevant in the following can be written as
\begin{equation}\label{eq:mono}
 \mathcal{U}(z,\G,\B,\Theta)=S_\G\,\mathcal{U}(z)\,S_\G^{-1}\,,\qquad\text{with}\qquad  S_\G=\prod_{k=1}^N S_{j_k}\,.
\end{equation} 
As follows from \eqref{eq:strans}, the similarity transform $S_\G$ generates the double-row monodromy $\mathcal{U}(z,\G,\B,\Theta)$ with single-row monodromies  $S_\G\mathcal{M}(z)S_\G^{-1}$ and $S_\G\hat{\mathcal{M}}(z)S_\G^{-1}$ which have Lax operators $\bar{\lax}_{j_k}$ at sites $j_k$ for $k=1,\ldots,N$ corresponding to the end points of the lines in $\G$, cf.~\eqref{eq:doublerow}. As the K-matrix only acts in the auxiliary space it remains unchanged. Furthermore, we identified the inhomogeneities of the double-row monodromy on the right hand side of \eqref{eq:mono} with the rapidities as 
\begin{equation}\label{eq:chainid}
{
\begin{array}{lll}
 v_{i_k}=\theta_k,& v_{j_k}=-\theta_k-1\quad\text{if}\qquad&k\in B\,,\\
&&\\
 v_{i_k}=\theta_k,&v_{j_k}=+\theta_k-1\quad\text{if}\qquad&k\in \bar B\,,
\end{array}      }
\end{equation}
where $k=1,\ldots,N$ and $\bar B$ denotes the complement of $B$, cf. \eqref{eq:B}.
Finally, we present the invariance condition satisfied by any lattice introduced in Section~\ref{sec:BBL}. It reads
\begin{equation}\label{eq:yinvcond}
 \mathcal{U}_\aux(z,\G,\B,\Theta)|\Psi_{\G}^{\B}(\Theta)\rangle=\Lambda(z,\G,\B,\Theta)\K(z)|\Psi_{\G}^{\B}(\Theta)\rangle\,.
\end{equation}
with
\begin{equation}\label{eq:lambda}
\Lambda(z,\G,\B,\Theta)= \prod_{k\in\B} f(z,\theta_k)\prod_{k\in\bar \B}f(z,-\theta_k)
\end{equation} 
and 
\begin{equation}\label{eq:blinefac}
f(z,\theta)=(z-\theta-1)(z-\theta+1)(z+\theta)(z+\theta+2) \,.
\end{equation} 

The function $\Lambda(z,\G,\B,\Theta)$ is derived toward the end of this section. The eigenvectors satisfying \eqref{eq:yinvcond} can be identified with the the partition function introduced in Section~\ref{sec:BBL} via
\begin{equation}\label{eq:rew}
 Z(\G,\B,\raps,\boldsymbol\alpha,\boldsymbol\beta)=\myfrac[1.5pt]{ \langle\boldsymbol\alpha,\boldsymbol\beta|\Psi_{\G}^{\B}(\Theta)\rangle}{\langle\boldsymbol\alpha_0,\boldsymbol\beta_0|\Psi_{\G}^{\B}(\Theta)\rangle}\,.
\end{equation} 
Here we used the notation $\langle\boldsymbol\alpha,\boldsymbol\beta|$ to denote the contraction with the $\langle \alpha_{k}|$ and $\langle \beta_{k}|$ at the corresponding sites $i_k$ and $j_k$ determined by the set $\G$ respectively.
To match with the conventions in Section~\ref{sec:BBL} we divided by the normalisation $\langle\boldsymbol\alpha_0,\boldsymbol\beta_0|\Psi_{\G}^{\B}(\Theta)\rangle$ which cannot be fixed by \eqref{eq:yinvcond} and in particular ensures \eqref{eq:defZ0}. A proof is provided in Section~\ref{ssec:init}.
Furthermore, we note that $\K$ is a diagonal matrix such that the condition \eqref{eq:yinvcond} implies that the off-diagonal entries in the auxiliary space of the double-row monodromy annihilate the state $|\Psi_{\G}^{\B}(\Theta)\rangle$. Thus after choosing an appropriate normalisation one finds that $|\Psi_{\G}^{\B}(\Theta)\rangle$ by the generators of the twisted Yangian obtained from the expansion the double-row monodromy in \eqref{eq:yinvcond}.

In the remaining part of this section we discuss two elementary sample invariants, study how to generate other invariants to show \eqref{eq:rew} and fix the function $\Lambda(z,\G,\B,\Theta)$.
\subsection{Line invariant}
The first invariant we encounter is the line solution in \eqref{eq:idmatrix} with 
\begin{equation}
  \G=((2,1)),\qquad \B=\emptyset,\qquad \bar\B=\{1\},\qquad\Theta=\{\theta_1\}\,.
\end{equation} 
From \eqref{eq:rew} we define the invariant corresponding to the line in \eqref{eq:idmatrix} up to a normalisation as the four-vector
\begin{equation}\label{eq:line}
|\Psi_{\cap}\rangle=|\Psi_{((2,1))}^\emptyset(\theta_1)\rangle=(1,0,0,1)\,.
 \end{equation} 
It is a solution to the Yangian invariance conditions in \eqref{eq:yinvcond}. This can be seen when considering the monodromy 
\begin{equation}\label{eq:linemon}
\mathcal{U}(z,((2,1)),\emptyset,\theta_1)=\bar{\lax}_{1}(z-\theta_1+1)\lax_{2}(z-\theta_{1})\K(z)\lax_{2}(z+\theta_{1})\bar{\lax}_{1}(z+\theta_1-1)\,,
\end{equation} 
cf.~\eqref{eq:mono}. First we note that the  unitarity relation in \eqref{eq:unitarity} is equivalent to the relations
\begin{equation}\label{eq:un1}
 \lax_{2}(z+\theta_{1})\bar{\lax}_{1}(z+\theta_1-1) |\Psi_{\cap}\rangle=(z+\theta_1-1)(z+\theta_1+1)|\Psi_{\cap}\rangle\,,
\end{equation} 
and 
\begin{equation}\label{eq:un2}
 \bar{\lax}_{1}(z-\theta_1+1)\lax_{2}(z-\theta_{1})|\Psi_{\cap}\rangle=(z-\theta_1)(z-\theta_1+2)|\Psi_{\cap}\rangle\,.
\end{equation} 
This follows from transposing \eqref{eq:un1} and \eqref{eq:un2} in the first and second space respectively and using the relation
 \begin{equation}
  \lax^t(z)=-\bar{\lax}(-z-1)\,.
 \end{equation} 
We obtain the eigenvalue
\begin{equation}
\Lambda(z,((2,1)),\emptyset,\theta_1)=f(z,-\theta_1)\,.
\end{equation} 

\subsection{Boundary-line invariant}
The next invariant we encounter is the boundary-line solution
\begin{equation}
  \G=((2,1)),\qquad \B=\{1\},\qquad \bar\B=\emptyset,\qquad\Theta=\{\theta_1\}\,,
\end{equation} 
 which depends on the boundary parameter $q$. Up to a normalisation it reads
\begin{equation}\label{eq:bounce}
|\Psi_\wedge(\theta_1)\rangle=|\Psi_{((2,1))}^{\{1\}}(\theta_1)\rangle=K_1(\theta_1)|\Psi_\cap\rangle=\frac{1}{q+\theta_1}(q+\theta_1,0,0,q-\theta_1)\,.
 \end{equation} 
 The corresponding double-row monodromy is the same as for the line solution, cf. \eqref{eq:linemon}, but the identification of the rapidities with the inhomogeneities changes according to \eqref{eq:chainid}. As a consequence of the boundary Yang-Baxter equation \eqref{eq:bybe} the boundary-line invariant in \eqref{eq:bounce} satisfies
 \begin{equation}\label{eq:bybevec}
\lax_{2}(z-\theta_{1})\K_a(z)\lax_{2}(z+\theta_{1})|\Psi_\wedge(\theta_1)\rangle_{1,2}=K_1(\theta_1)\lax_{2}(z+\theta_{1})\K_a(z)\lax_{2}(z-\theta_{1}) |\Psi_\cap\rangle_{1,2}\,.
 \end{equation} 
Using this relation as well as the unitarity relations for the line solution in \eqref{eq:un1} and  \eqref{eq:un2} we find that
\begin{equation}
\Lambda(z,((2,1)),\{1\},\theta_1)=f(z,\theta_1)\,.
\end{equation} 
The sign of the rapidity of the function $f$ is changed. This is a consequence of the change of sign in the rapidity when bouncing off the boundary as observed in \eqref{eq:bybevec}. 

\subsection{Initial conditions and further invariants}\label{ssec:init}
So far we have shown that lines and boundary-lines are Yangian invariants. We can now combine line and boundary-line solutions to obtain Yangian invariants of monodromies of higher length, i.e. configurations
\begin{equation}\label{eq:ini}
  \G_0=((L,L-1),(L-2,L-3),\ldots,(2,1))\,,
\end{equation} 
with arbitrary $\B$. The corresponding invariant reads
\begin{equation}\label{eq:initial}
|\Psi_{\G_0}^\B(\Theta)\rangle=|\Psi_{\{\wedge,\cap\}}(\theta_1)\rangle\otimes\ldots\otimes|\Psi_{\{\wedge,\cap\}}(\theta_N)\rangle \,,
\end{equation} 
where depending on the set $\B$ the $|\Psi_{\{\wedge,\cap\}}(\theta)\rangle$ denotes the line invariant \eqref{eq:line} or the boundary-line invariant \eqref{eq:bounce}. 
See Figure~\ref{fig:init8} for an example wit $N=4$ lines.
As a consequence of the previous analysis of the two site invariants this configuration satisfies the Yangian invariance condition \eqref{eq:yinvcond} with
\begin{equation}\label{eq:fnull}
\Lambda(z,\G_0,\B,\Theta)= \prod_{k\in\B} f(z,\theta_k)\prod_{k\in\bar \B}f(z,-\theta_k)\,.
\end{equation} 

\begin{figure}[ht!]
\begin{center}
  \includegraphics[width=0.8\linewidth]{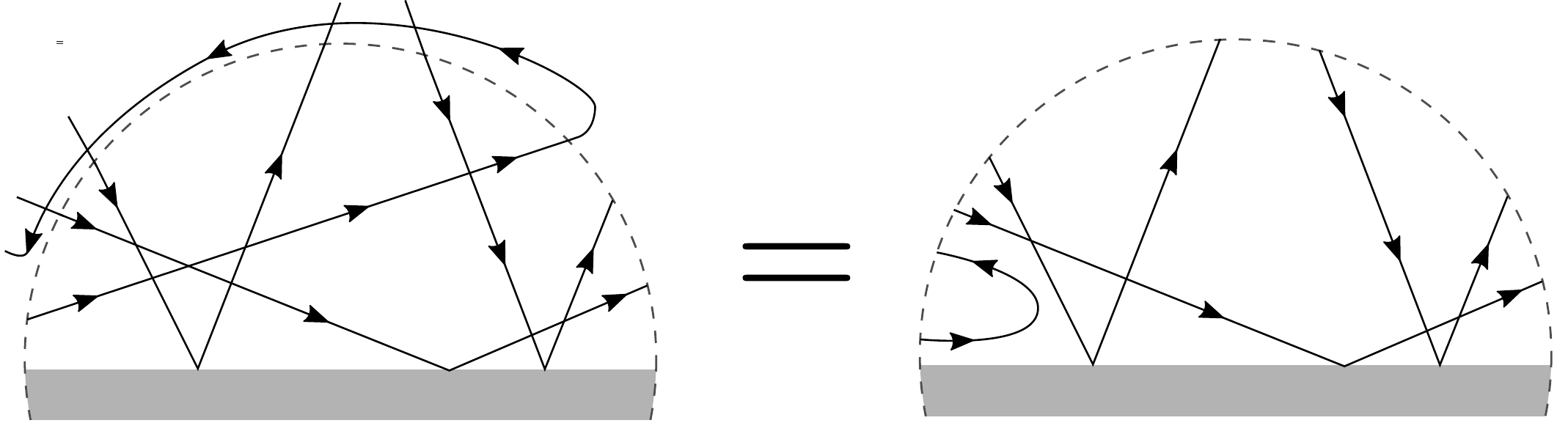}
\end{center}
  \caption{Schematic diagram of the recursion procedure.}
  \label{fig:recurs}
\end{figure}

By construction, every invariant can be related to the initial conditian \eqref{eq:initial} by a product of R-matrices, for a similar argumentation see \cite{Baxter1987}. Starting from an arbitrary configuration $G$ we note that for any lattice we have $i_1=L$. Then the index $i_2$ can take the values $i_2=L-1,L-2$. If $i_2=L-2$ we have $j_1=L-1$ and nothing to do, cf.~\eqref{eq:ini}. Else $i_2=L-1$ and $L-2\geq j_1\geq 1$. In this case we use the relation
\begin{equation}\label{eq:recu}
\begin{split}
|\Psi_{\G'}^\B(\Theta)\rangle&= \mathbb{P}_{j_1-1j_1}\cdots \mathbb{P}_{L-1,j_1} S_G  R_{L-1,j_1}R_{L-2,j_1}\cdots R_{j_1-1j_1}S_G^{-1}|\Psi_{\G}^\B(\Theta)\rangle\,,
\end{split}
\end{equation} 
which relates the configuration labelled by the set $\G=((L,j_1),(L-2,j_2),\ldots,(i_N,j_N))$
to $\G'=((L,L-1),(L-2,j_2'),\ldots,(i_N',j_N'))$. Here $R_{ij}$ denotes the R-matrix $R_{ij}=R_{ij}(v_i-v_j)$ with the inhomogeneities identified according to \eqref{eq:chainid} depending on the set $\B$. The matrix $\mathbb{P}$ denotes the permutation acting on a tensor product of two vectors $v,w$ as $\mathbb{P}(v\otimes w)=w\otimes v$. See Figure~\ref{fig:recurs} for a diagrammatic portrayal of \eqref{eq:recu}.

Proceeding recursively we obtain a relation among any lattice configuration and the initial one \eqref{eq:initial}. Diagrammatically it is straightforward to show that
\begin{equation}\label{eq:exch}
\begin{split}
\mathcal{U}(z,G',B,\Theta) &\mathbb{P}_{j_1-1j_1}\cdots \mathbb{P}_{L-1,j_1} S_G  R_{L-1,j_1}R_{L-2,j_1}\cdots R_{j_1-1j_1}S_G^{-1}\\&= \mathbb{P}_{j_1-1j_1}\cdots \mathbb{P}_{2j_1} S_G  R_{L-1,j_1}R_{L-2,j_1}\cdots R_{j_1-1j_1}S_G^{-1}\mathcal{U}(z,G,B,\Theta)
\end{split}
\end{equation} 
and thus we obtain
\begin{equation}
\begin{split}
\mathcal{U}(z,G',B,\Theta)|\Psi_{\G'}^\B(\Theta)\rangle&= \mathbb{P}_{j_1-1,j_1}\cdots \mathbb{P}_{L-1,j_1}\\&\quad\times S_G  R_{L-1,j_1}R_{L-2,j_1}\cdots R_{j_1-1,j_1}S_G^{-1}\mathcal{U}(z,G,B,\Theta)|\Psi_{\G}^\B(\Theta)\rangle\\
\end{split}
\end{equation} 

This procedure is invertible and allows us to relate a lattice of given $G$ to the initial condition \eqref{eq:initial}. As a consequence, we find that every Baxter lattice with given sets $\G$ and $\B$ satisfies the Yangian invariance condition \eqref{eq:yinvcond} with 
\begin{equation}
 \Lambda(z,\G,\B,\Theta)=\Lambda(z,\G_0,\B,\Theta)\,,
\end{equation} 
as given in \eqref{eq:fnull}.

\section{Algebraic Bethe ansatz for invariants of the twisted Yangian}\label{sec:ABA}
In the following we construct invariants of the twisted Yangian satisfying the condition \eqref{eq:yinvcond} using the algebraic Bethe ansatz for integrable models with boundary \cite{Sklyanin:1988yz}. We find that the Bethe ansatz equation are exactly solvable  and present the corresponding Bethe roots and Baxter Q-functions. This allows us the obtain another representation of the partition function \eqref{eq:defZ} in terms of creation operators on a reference state.

It is convenient to define the $2^L\times2^L$ matrices  $\mathcal{A}(z)$, $\mathcal{B}(z)$, $\mathcal{C}(z)$ and $\mathcal{D}(z)$ as the entries of the double-row monodromy matrix \eqref{eq:mono} in the auxiliary space
\begin{equation}
 \mathcal{U}_\aux(z,\G,\B,\Theta)=\left(\begin{array}{cc}
                       \mathcal{A}(z)&\mathcal{B}(z)\\
                       \mathcal{C}(z)&\mathcal{D}(z)
                      \end{array}\right)\,.
\end{equation} 
Here we suppressed the dependence on the rapidities. Following \cite{Sklyanin:1988yz} we introduce the operator 
\begin{equation}
\mathcal{D}(z)= \tilde{\mathcal{D}}(z)+\frac{1}{2z+1}\mathcal{A}(z)\,,
\end{equation} 
in order to diagonalise the transfer matrix. The fundamental commutation relations which arise from the boundary Yang-Baxter equation \eqref{eq:ubybe} and are relevant in the following can be found in Appendix~\ref{app:openfcr}. The ansatz for the eigenvectors of the transfer matrix reads
\begin{equation}\label{eq:baestate}
 |\psi_m\rangle=\mathcal{B}(z_1)\cdots \mathcal{B}(z_m)|\Omega\rangle\,,
\end{equation} 
where $|\Omega\rangle=S_\G\,|1\rangle\otimes\ldots\otimes|1\rangle$ denotes a reference state with
\begin{equation}
\ua(z)|\Omega\rangle= \alpha(z)|\Omega\rangle\,,\qquad  \udt(z)|\Omega\rangle= \tilde\delta(z)|\Omega\rangle\,,\qquad \uc(z)|\Omega\rangle=0\,.
\end{equation} 
The functions $\alpha(z)$ and $\tilde\delta(z)$ can be determined from the definition of the double-row monodromy \eqref{eq:mono}. By acting on the reference state one obtains 
\begin{equation}
 \alpha(z)=(q+z)\Xi(z,\G,\B,\Theta)\,,\qquad \tilde\delta(z)=\frac{2z}{2z+1}(q-z-1)\Xi(z-1,\G,\B,\Theta)\,.
\end{equation}
Here the function $\Xi$ only depends on the set $\B$ and $\bar \B$ denoting the reflected and unreflected lines respectively. We have
\begin{equation}\label{eq:xi}
\Xi(z,\G,\B,\Theta)=\prod_{k\in \B}g(z,\theta_k)\prod_{k\in\bar \B} g(z,-\theta_k)\,,
\end{equation} 
with
\begin{equation}
g(z,\theta_k)= (z-\theta_k)(z-\theta_k+1)(z+\theta_k+1)(z+\theta_k+2)\,.
\end{equation} 
From the fundamental commutation relations in Appendix~\ref{app:openfcr} we obtain the action of the diagonal elements of the double-row monodromy on our off-shell Bethe states \eqref{eq:baestate}. It reads
\begin{equation}\label{eq:uwa}
 \mathcal{A}(z)|\psi_m\rangle=\alpha(z)\frac{Q(z-1)}{Q(z)}|\psi_m\rangle+\sum_{k=1}^m M_k(z,z_i)|\psi^{(k)}_m\rangle\,,
\end{equation} 
and
\begin{equation}\label{eq:uwd}
 \tilde{\mathcal{D}}(z)|\psi_m\rangle=\tilde\delta(z)\frac{Q(z+1)}{Q(z)}|\psi_m\rangle+\sum_{k=1}^m N_k(z,z_i)|\psi^{(k)}_m\rangle\,,
\end{equation} 
with $|\psi^{(k)}_m\rangle= \mathcal{B}(z_1)\cdots \mathcal{B}(z_{k-1})\mathcal{B}(z) \mathcal{B}(z_{k+1})\cdots \mathcal{B}(z_{m})|\Omega\rangle$ and Baxter's Q-function
\begin{equation}
 Q(z)=\prod_{i=1}^m (z-z_i)(z+z_i+1)\,.
\end{equation} 
Here the coefficients $M_k$ and $N_k$ proportional to $|\psi_m^{(k)}\rangle$ are usually referred to as the ``unwanted terms'' and can be found in \eqref{eq:mk} and \eqref{eq:nk} respectively.
While in the ordinary Bethe ansatz for the transfer matrix one demands that the sum of those terms vanishes we want them to vanish individually. This would yield a common eigenstate of $\mathcal{A}(z)$ and $\mathcal{D}(z)$. In order to determine the Bethe roots for which \eqref{eq:baestate} is a solution of the Yangian invariance condition \eqref{eq:yinvcond} we first make sure that the eigenvalues match. From \eqref{eq:uwa} and \eqref{eq:uwd} we obtain the conditions
\begin{equation}\label{eq:q1}
\Xi(z,\G,\B,\Theta)\frac{Q(z-1)}{Q(z)}=\Lambda(z,\G,\B,\Theta)\,,
\end{equation} 
and 
\begin{equation}\label{eq:q2}
\Xi(z-1,\G,\B,\Theta)\frac{Q(z+1)}{Q(z)}=\Lambda(z,\G,\B,\Theta)\,.
\end{equation} 
This is the counterpart of the degenerate Baxter equation obtained in \cite{Frassek2014} for closed spin chains. By construction we already know the explicit expressions of $\Lambda$ and $\Xi$, cf.~\eqref{eq:lambda} and \eqref{eq:xi}. They satisfy the identity
\begin{equation}
\frac{ \Lambda(z+1,\G,\B,\Theta)}{\Xi(z+1,\G,\B,\Theta)}=\frac{ \Xi(z-1,\G,\B,\Theta)}{\Lambda(z,\G,\B,\Theta)}\,.
\end{equation} 
Thus after shifting the spectral parameter $z$ we find that \eqref{eq:q1} and \eqref{eq:q2} are equivalent. The remaining equation is exactly solved by the Q-function
\begin{equation}
 Q(z)=\prod_{k\in\B} (z-\theta_k)(z+\theta_k+1)\prod_{k\in\bar \B}(z+\theta_k)(z-\theta_k+1)\,.
\end{equation}
The magnon number is thus given by $m=N$.
Note that the Q-function is symmetric under the transform $z\rightarrow -z-1$. We can read off the Bethe roots
\begin{equation}\label{eq:betheroots}
{
\begin{array}{lllll}
 z_{k}=+\theta_k&\vee & z_{k}=-\theta_k-1&\text{if}&k\in \B\,,\\
&\\
 z_{k}=-\theta_k&\vee & z_{k}=+\theta_k-1&\text{if}&k\in\bar \B\,.\\
\end{array}      }
\end{equation}  
We further note that the creation operator satisfies
\begin{equation}
 \mathcal{B}(z)=-\frac{z}{z+1}\mathcal{B}(-z-1)\,,
\end{equation} 
cf. \eqref{eq:BasB}, thus the Bethe vector \eqref{eq:baestate} is invariant  up to a factor under the transformation $z_i\rightarrow -z_i-1$. We conclude that there is exactly one solution to the Bethe equations and therefore one eigenvector for a given lattice labelled by $G$ and $B$. 

We arrive at the expression of the partition function \eqref{eq:defZ} written in terms of the creation operators $\mathcal{B}$ acting on the reference state $|\Omega\rangle$ in the framework of the algebraic Bethe ansatz
\begin{equation}\label{eq:abaZ}
 Z(\G,\B,\raps,\boldsymbol\alpha,\boldsymbol\beta)=\myfrac[1.5pt]{\langle\boldsymbol\alpha,\boldsymbol\beta|\mathcal{B}(z_1)\cdots \mathcal{B}(z_N)|\Omega\rangle}{\langle\boldsymbol\alpha_0,\boldsymbol\beta_0|\mathcal{B}(z_1)\cdots \mathcal{B}(z_N)|\Omega\rangle}\,.
\end{equation} 
Here the Bethe roots have to be identified appropriately via \eqref{eq:betheroots}. 
Note that due to the commutation relations \eqref{eq:opB} the off-shell Bethe vector \eqref{eq:baestate} is symmetric in the Bethe roots $z_i$.

In the following section we show that \eqref{eq:abaZ} indeed yields the partition function and thus conclude that the constructed Bethe vectors not only are eigenstates of $\ua$ and $\ud$ but also are annihilated by $\ub$ and $\uc$.

\subsection{Initial conditions from ABA}
We have seen that the action of the operators $\ua$ and $\ud$ yields the desired eigenvalues under the identification of the Bethe roots \eqref{eq:betheroots}. We now argue that also $\mathcal{B}$ and $\mathcal{C}$ vanish by evaluating the partition function for the initial condition in analogy to \cite{Baxter1987}.
\begin{figure}[ht!]
\begin{center}
  \includegraphics[width=0.5\linewidth]{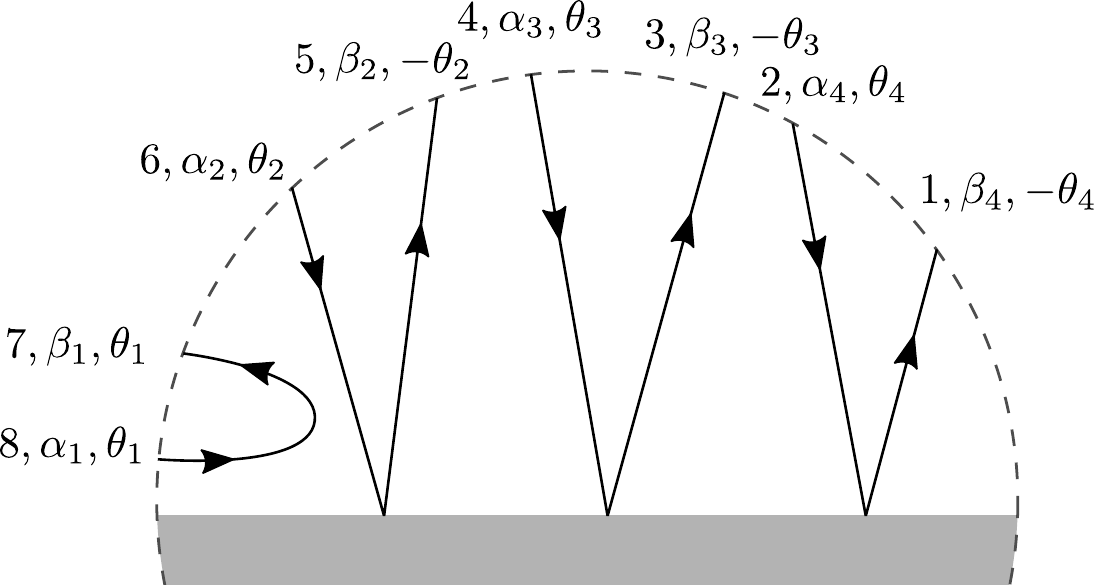}
\end{center}
  \caption{Initial condition $G=((8,7),(6,5),(4,3),(2,1))$ with $B=\{2,3,4\}$.}
  \label{fig:init8}
\end{figure}

For a generic lattice the initial configuration is given by 
\begin{equation}
 G=((2N,2N-1),(2N-2,2N-3),\ldots,(2,1))\,,
\end{equation} 
with some set of boundary lines $B$, cf.~Figure~\ref{fig:init8}. As discussed in in the previous section we have to identify
\begin{equation}\label{eq:chainidinit}
{
\begin{array}{lll}
v_{2(N-k+1)}=\theta_k,& v_{2(N-k)+1}=-\theta_k-1\quad\text{if}\qquad&k\in B\,,\\
&&\\
 v_{2(N-k+1)}=\theta_k,&v_{2(N-k)+1}=+\theta_k-1\quad\text{if}\qquad&k\in \bar B\,.
\end{array}      }
\end{equation}
cf. \eqref{eq:chainid}, to obtain the corresponding monodromy. 

\begin{figure}[ht!]
\begin{center}
  \includegraphics[width=0.4\linewidth]{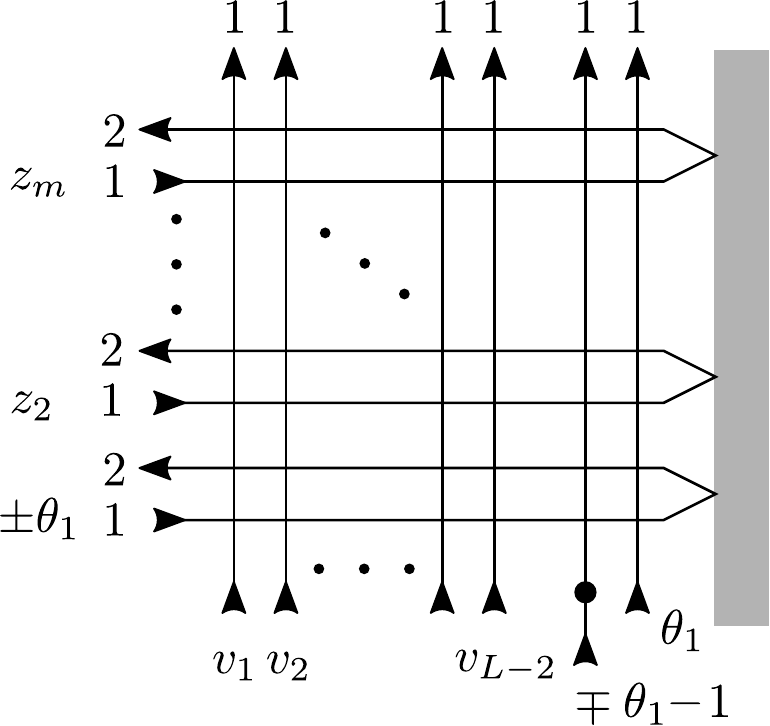}
\end{center}
  \caption{Vertex representation of the Bethe vector with length $2N$ and magnon number $m$. Here the similarity transform on site $L-1$ is denoted by a black dot. The inhomogeneities and Bethe roots are identified as in \eqref{eq:ident}.}
  \label{fig:Bvec}
\end{figure}

\begin{figure}[ht!]
\begin{center}
  \includegraphics[width=0.6\linewidth]{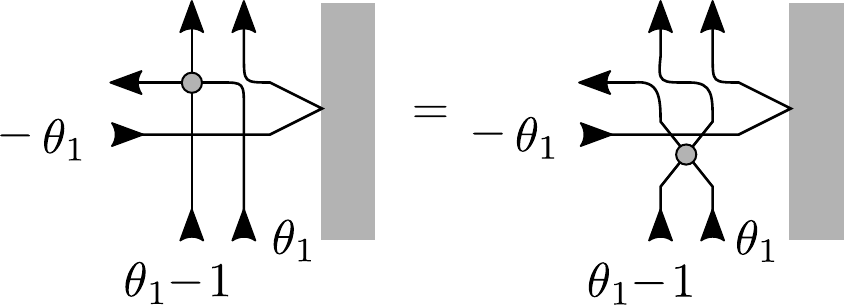}
\end{center}
  \caption{Element with $v_{2N}=\theta_1,\, v_{2N-1}=\theta_1-1$ and $z=-\theta_1$. The dotted vertex denotes the antisymmetric projector $\mathcal{L}(-1)$.}
  \label{fig:Bvecss}
\end{figure}

\begin{figure}[ht!]
\begin{center}
  \includegraphics[width=0.6\linewidth]{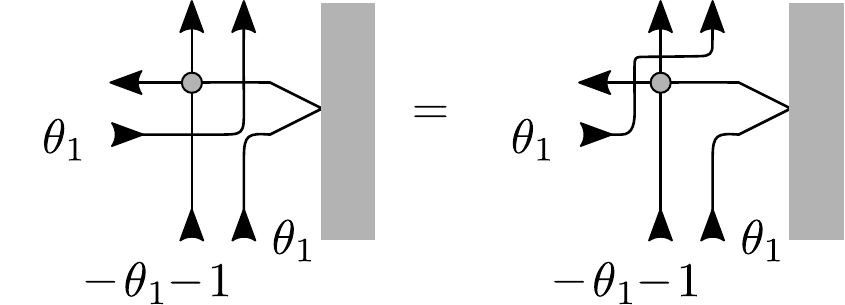}
\end{center}
  \caption{Element with $v_{2N}=\theta_1,\, v_{2N-1}=-\theta_1-1$ and $z=\theta_1$. Again, the dotted vertex denotes the antisymmetric projector $\mathcal{L}(-1)$.}
  \label{fig:Bvecs}
\end{figure}

To show that under the identification \eqref{eq:chainidinit} the Bethe vector defined in \eqref{eq:baestate} yields the initial condition with the choice of Bethe roots as stated in \eqref{eq:betheroots} we consider a general off-shell Bethe state of length $2N$ and magnon number $m$
\begin{equation}
 |\psi_{2N,m}\rangle=\mathcal{B}(z_1)\cdots\mathcal{B}(z_m)|\Omega\rangle
\end{equation} 
Here we explicitly spelled out the dependence on the length of the spin chain. Now, we identify only one of the Bethe roots and two of the rapidities and leave the rest arbitrary 
\begin{equation}\label{eq:ident}
 v_{2N}=\theta_1,\qquad v_{2N-1}=\mp\theta_1-1,\qquad z_1=\pm\theta_1\,.
\end{equation} 
Under this identification we can express $|\psi_{2N,m}\rangle$ in terms of a Bethe state of lower length and magnon number $|\psi_{2N-2,m-1}\rangle$. A convenient way to show this relation is diagrammatically.
In general a Bethe vector can itself be represented as a lattice as shown in Figure~\ref{fig:Bvec}. 
We proceed in analogy to  \cite{Baxter1987} in the framework of the coordinate Bethe ansatz.
In the algebraic framework it is important to notice that given the identification \eqref{eq:ident} two Lax matrices in the $\mathcal{B}$ operators are evaluated at the special values 
\begin{equation}\label{eq:specp}
\mathcal{L}(0)=\mathbb{P}, \qquad  \mathcal{L}(-1)=-2\mathbb{A}, 
\end{equation} 
where $\mathbb{P}=\sum_{a,b=1}^2e_{ab}\otimes e_{ba}$ denotes the permutation operator and $\mathbb{A}=\frac{1}{2}(\mathbb{I}-\mathbb{P})=Y\dot Y^t$ the antisymmetric projector on the singlet with $Y^t=(0,+1,-1,0)$.
First we argue that the Bethe vector evaluates to zero if $\alpha_1\neq \beta_1$, cf. Figure~\ref{fig:Bvec}. This can be seen from the matrixelemet in the lower right corner. Using \eqref{eq:specp} and the Yang-Baxter equation we can make the antisymmetric projector act on the external states. This is shown in Figure~\ref{fig:Bvecss} and \ref{fig:Bvecs} where the antisymmetric projector is denoted by the dotted vertex and the dotted line denotes the matrix $S$ in \eqref{eq:strans}. We find that the Bethe vector vanishes for $\alpha_1\neq \beta_1$.

We now focus on the case where $\alpha_1=\beta_1$. Here it is convenient to take $z_m=\pm\theta_1$. This can be done as the Bethe vector is symmetric in the Bethe roots, cf.~\eqref{eq:opB}. This time we have the matrix elements depicted in Figure~\ref{fig:Bvecss} and \ref{fig:Bvecs} in the upper right corner of Figure~\ref{fig:Bvec}. 
Noting that the vertices are frozen in the upper two rows and using the bootstrap equations
\begin{equation}
 (\mathcal{L}_a(z)\mathcal{L}_b(z-1))Y_{ab}=(z+1)(z-1)\mathbb{I}\otimes Y_{ab}\,,
\end{equation} 
and
\begin{equation}
 (\mathcal{L}_b(z)\mathcal{L}_a(z-1))Y_{ab}=(z+1)(z-1)\mathbb{I}\otimes Y_{ab}\,,
\end{equation} 
we can evaluate the Bethe vector and relate it to the Bethe state of lower length. Under the identification in \eqref{eq:ident} we obtain
\begin{equation}
|\psi_{2N,m}\rangle=|\psi_{2N-2,m-1}\rangle\otimes|\phi\rangle \,,
\end{equation} 
with
\begin{equation}
|\phi\rangle= \begin{cases} h(+\theta_1)|\Psi_{\wedge}(\theta_1)\rangle &\mbox{if } 1\in \B \\
h(-\theta_1)|\Psi_{\cap}(\theta_1)\rangle & \mbox{if } 1\notin \B\end{cases}\,.
\end{equation}
Here we defined the function
\begin{equation}
\begin{split}
 h(\theta_1)=4\theta_1(\theta_1+1)(q+\theta_1)&\prod_{i=2}^N (\theta_1+z_i+2) (\theta_1-z_i-1) (\theta_1+z_i) (\theta_1-z_i+1)\\&\times\prod_{k=1}^{2N-2}  (\theta_1+v_k)(\theta_1-v_k+1)\,.  
     \end{split}
\end{equation} 
Proceeding recursively we arrive at the invariant for the initial condition \eqref{eq:initial} up to a normalisation. As discussed this normalisation cancels in the definition of the partition function \eqref{eq:abaZ}. We have thus shown that \eqref{eq:abaZ} yields the correct partition function \eqref{eq:defZ} for the initial configuration. Also as shown in Section~\ref{ssec:init} the invariant for the initial condition satisfies the Yangian invariance condition \eqref{eq:yinvcond}. Thus $\mathcal{B}$ and $\mathcal{C}$ vanish on the corresponding Bethe vector evaluated here.

Following the same logic as in Section~\ref{ssec:init} we find that any other Bethe vector with the choice of rapidities in \eqref{eq:chainid} and Bethe roots \eqref{eq:betheroots} satisfies the Yangian invariance condition. This follows from the exchange properties given in \eqref{eq:exch} which hold for every element of the double-row monodromy and in particular the $\mathcal{B}$-operators. 
\section{Invariants from the coordinate wave function }\label{sec:CBA}
Finally, we would like to present the partition function expressed in terms of the coordinate wave function of the open XXX spin chain. Such a formula can be obtained from the expression derived using the algebraic Bethe ansatz \eqref{eq:abaZ} when expressing the creation operators of the open chain in terms of the ones of closed chains, see e.g.~\cite{Wang2002,Kitanine:2007bi}, and using its well known coordinate wave function.

First we note that the two single-row monodromies in the double row monodromy are not independent. They can be related via
\begin{equation}\label{eq:singmon}
 \hat{{\mathcal{M}}}^{t_a}(z)=(-1)^LS{\mathcal{M}}(-z-1)S^{-1}\,,
\end{equation} 
where $t_a$ denotes the transposition in the auxiliary space and $S$ was defined in \eqref{eq:strans}. As a consequence we can express the creation operator $\mathcal{B}$ of the open chain in terms of  the single-row monodromy matrix in the auxiliary space 
\begin{equation}\label{eq:closmon}
S_G{\mathcal{M}}(z)S_G^{-1}=\left(\begin{array}{cc}
                             A(z)&B(z)\\
                             C(z)&D(z)
                            \end{array}\right)\,.
\end{equation} 
One finds
\begin{equation}\label{eq:BasB}
\begin{split}
 \mathcal{B}(z)
 &=(-1)^{L}\frac{2z}{2z+1}\sum_{\tau\in R_1}(-1)^{|\tau|}(q-z^\tau-1) B(z^\tau)A(-z^\tau-1)\,,
 \end{split}
\end{equation} 
where $\tau$ denotes the reflections in $R_1$ given by $z\rightarrow -z-1$ and $|\tau|$ counts the number of such reflections. From the definition of the double-row monodromy in \eqref{eq:doublerow} and using the fundamental commutation relations for the closed chain in Appendix~\ref{app:closed} we can write the off-shell Bethe vector introduced in \eqref{eq:baestate} as
\begin{equation}\label{eq:vecmitb}
 |\psi_m\rangle =\mathcal{N}_{L,m}\sum_{\tau\in R_m}(-1)^{|\tau|}\prod_{i<j}h(z_i^\tau,-z_j^\tau-1)\prod_{i=1}^m(q-z_i^\tau-1)\kappa(-z_i^\tau-1)B(z_i^\tau)|\Omega\rangle\,.
\end{equation} 
where the first sum denotes the sum over the $2^m$ reflections $R_m$ of the Bethe roots $z_i\rightarrow -z_i-1$ and  $\mathcal{N}_{L,m}=(-1)^{mL}\prod_{i=1}^m \frac{2z_i}{2z_i+1}$. The functions defined read
\begin{equation}
 \kappa(z)=\prod_{i=1}^L(z-v_i+1)\,,\qquad  h(x,y)=\frac{1+x-y}{x-y}\,.
\end{equation} 
The above relation \eqref{eq:vecmitb} can be derived using the commutation relations in Appendix~\ref{app:closed} and the identity
\begin{equation}
 \sum_{\tau\in R_2} (-1)^{|\tau|} (q-z_i^\tau-1)(q-z^\tau_j-1)k(z^\tau_i,-z^\tau_j-1)=0\,,
\end{equation} 
with $k(x,y)=1/(x-y)$, cf.~\eqref{eq:closedcom}.
Now, it is straight forward to obtain the coordinate wave function of the open spin chain introduced in Section~\ref{sec:twistedY} as the translation among these two representations is well understood for the closed chain, see e.g.~\cite{Essler2005}. The off-shell Bethe vectors in the coordinate representation can be found in Appendix~\ref{app:closedwave}. We find that the off-shell Bethe state in \eqref{eq:baestate} can be written as
\begin{equation}
|\psi_m\rangle =\mathcal{N}_{L,m}\sum_{1\leq x_1<\ldots<x_m\leq L}\Upsilon( \mathbf{v}, \mathbf{z}, \mathbf{x})S_G\,e_{21}^{(x_1)}\cdots e_{21}^{(x_m)} S_G^{-1}|\Omega\rangle\,,
\end{equation} 
where the parameters $x_i\in \mathbf{x}$ with $1\leq x_i\leq L$ denote the positions of the magnons on the chain, $\mathbf{z}$ the set of Bethe roots $z_i$ and $\mathbf{v}$ the set of inhomogeneities $v_i$. The coordinate wave function takes the form
\begin{equation}\label{eq:wave}
 \Upsilon( \mathbf{v}, \mathbf{z}, \mathbf{x})=\sum_{\tau\in R_m}\sum_{\sigma\in S_m}(-1)^{|\tau|}{\rm A}(z_{\sigma(1)}^\tau,\ldots,z_{\sigma(m)}^\tau)\prod_{i=1}^m\phi_{x_i}(z_{\sigma(i)}^\tau, \mathbf{v})\,.
\end{equation} 
Here the amplitude part reads
\begin{equation}
 {\rm A}(z_{\sigma(1)},\ldots,z_{\sigma(m)})=\prod_{1\leq k<l\leq m}\frac{(z_{\sigma(k)}-z_{\sigma(l)}+1)(z_{\sigma(k)}+z_{\sigma(l)}+2)}{(z_{\sigma(k)}-z_{\sigma(l)})(z_{\sigma(k)}+z_{\sigma(l)}+1)}\,,
\end{equation} 
and the wave part takes the form
\begin{equation}
 \phi_{x}(z, \mathbf{v})=(-1)^L (q-z-1)\prod_{j=1}^L(z+v_j)\prod_{j=1}^{x-1}(z-v_{j}+1)\prod_{j=x+1}^L(z-v_{j})\,.
\end{equation} 
The sum in the wave function \eqref{eq:wave} goes over $m!$ permutations $S_m$ and $2^m$ reflections $R_m$, i.e.
\begin{equation}
 z_i\rightarrow -z_i-1 \quad\text{for} \quad i=1,\ldots,m\,.
\end{equation} 

Taking $L=2N$ and $m=N$ we obtain the Bethe vectors relevant to calculate the partition function \eqref{eq:defZ}. The positions of the magnons $x_i$ follow from the set $\G$ and the external states at the perimeter $\boldsymbol\alpha$ and $\boldsymbol\beta$. As discussed in \cite{Frassek2014} we have the following rule to determine the set of magnon numbers from the state configuration at the perimeter
\begin{equation}
\mathbf{x}=\{i_k|\alpha_k=2\}\cup \{j_k|\beta_{k}=1\}\,,\qquad k=1,\ldots,N\,.
\end{equation} 
Identifying the inhomogeneities and the Bethe roots with the rapidities as discussed in the previous section we arrive at the expression for the partition function using \eqref{eq:abaZ}. It can be written as
\begin{equation}\label{eq:cbaZ}
 Z(\G,\B,\raps,\boldsymbol\alpha,\boldsymbol\beta)=(-1)^{\mathcal{I}(\boldsymbol\beta)}\myfrac[1.5pt]{\Upsilon(\G,\B,\Theta, \mathbf{x})}{\Upsilon(\G,\B,\Theta, \mathbf{x}_0)}\,,
\end{equation} 
where the function $\mathcal{I}(\boldsymbol\beta)$ counts the total number of states $\beta_i=2$ and is an artefact of the transformation matrix $S_G$. 

\section{Conclusions}
We defined Baxter lattices with boundary and studied their partition function. In particular we gave a recipe to evaluate the partition function from the off-shell Bethe states in the coordinate and algebraic Bethe ansatz framework. Furthermore, we argued that the resulting expressions can be interpreted as invariants of the twisted Yangian generated by the double-row monodromy. 

We found that all invariants of the Yangian of the {\small XXX} spin chain monodromy which can be obtained from \cite{Baxter1987,Frassek2014}  are also invariants of the twisted Yangian considered here.
It should be straightforward to lift the results obtained to the trigonometric case where the spin chain would be given by the open {\small XXZ} spin chain. The generalisation to higher rank algebras and in particular using realisations along the lines of \cite{Frassek2014} and \cite{Chicherin2014b}, see also \cite{Kanning:2014cca} where a relation to matrix models is discussed, would be interesting.
In the case of the twisted Yangian there might be a similar story for the relation between the Yangian and the Gra\ss mannian, see e.g. \cite{Arkani-Hamed2012}.
A more challenging task is to generalise our construction to the case with non-diagonal boundary where the off-diagonal Bethe ansatz \cite{offdiag} applies. For this purpose it might in particular be interesting to find a purely algebraic prove that the eigenvectors of the operators $\ua$ and $\ud$ are annihilated by $\ub$ and $\uc$ as presented for the single-row monodromy in \cite{Frassek2014}.

The expressions for the partition function in terms of the Bethe vectors derived in this article can be used to compute the partition functions with domain wall boundary conditions. 
It might be interesting to rederive the determinant formula for the six-vertex model with reflecting end \cite{Tsuchiya1998a} and relate it to the approach in \cite{Gier2011,Galleas2014}.

Another interesting direction to pursue is to understand whether the Q-functions obtained here explicitly have an interpretation as spectral determinants as studied in the continuum limit of the {\small XXZ} chain in the {\small ODE/IM} correspondence \cite{Dorey1999,Bazhanov2003}. Here the Q-operators for the open Heisenberg chain were recently constructed in \cite{Frassek:2015mra} may become practical.

The work \cite{Frassek2014} was predominantly motivated by the study of scattering amplitudes in planar $\mathcal{N}=4$ super Yang-Mills theory where the Yangian symmetry becomes manifest when formulated in terms of on-shell diagrams \cite{Arkani-Hamed2012,Arkani-Hamed2015} in the planar \cite{Drummond:2010uq} but also non-planar \cite{Frassek2016} contribution to amplitudes. It remains to see whether the invariants studied here will reappear in the study of integrability in the {\small AdS/CFT} correspondence where boundary and related problems where studied in e.g.~\cite{Berenstein:2005vf,Drukker:2012de,deLeeuw:2015hxa}. 

\section*{Acknowledgements}
I like to thank Patrick Dorey, Nadav Drukker, Paul Heslop, Nils Kanning, David Meidinger, Alejandro de la Rosa Gomez and Istvan Szecsenyi for interesting discussions.
I acknowledge the support of IH\'{E}S. 
A significant part of this research was carried out at the Department of Mathematical Sciences, Durham University (UK).
The research leading to these results has received funding from the People Programme
(Marie Curie Actions) of the European Union’s Seventh Framework Programme FP7/2007-
2013/ under REA Grant Agreement No 317089 (GATIS).
    \renewcommand*\appendixpagename{\Large Appendices}
\begin{appendices}
\addtocontents{toc}{\protect\setcounter{tocdepth}{0}}
\section{FCR: Open chain}\label{app:openfcr}
The relevant fundamental commutation relations that arise from the boundary Yang-Baxter equation \eqref{eq:ubybe} read
\begin{equation}\label{eq:opB}
 [\ub(x),\ub(y)]=0\,,
\end{equation} 
\begin{equation}
 \ua(x)\ub(y)=h_\ua(x,y)\ub(y)\ua(x)+g_\ua(x,y)\ub(x)\ua(y)+g_\udt(x,y)\ub(x)\udt(y)\,,
\end{equation} 
\begin{equation}
 \udt(x)\ub(y)=h_\udt(x,y)\ub(y)\udt(x)+k_\ua(x,y)\ub(x)\ua(y)+k_\udt(x,y)\ub(x)\udt(y)\,,
\end{equation} 
with
\begin{equation}
 \begin{array}{lll}
  h_\ua(x,y)=\frac{(x+y)(x-y-1)}{(x-y)(x+y+1)}\,,&  g_\ua(x,y)=\frac{2y}{(x-y)(2y+1)}\,,&  g_\udt(x,y)=\frac{-1}{x+y+1}\,,\\
  &&\\
  h_\udt(x,y)=\frac{(x-y+1)(x+y+2)}{(x-y)(x+y+1)}\,,&  k_\ua(x,y)=\frac{4y(x+1)}{(2x+1)(2y+1)(x+y+1)}\,,&  k_\udt(x,y)=\frac{-2(x+1)}{(x-y)(2x+1)}\,.\\
 \end{array}
\end{equation} 
The unwanted terms in \eqref{eq:uwa} and \eqref{eq:uwd} read
\begin{equation}\label{eq:mk}
\begin{split}
 M_k&(z,\mathbf{z})=g_\mathcal{A}(z,z_k)\prod_{i\neq k}h_\mathcal{A}(z_k,z_i)\alpha(z_k)+g_\udt(z,z_k)\prod_{i\neq k}h_\udt(z_k,z_i)\tilde\delta(z_k)\,\\
 &=-\left[\frac{\alpha(z_k)Q(z_k-1)}{(z-z_k)(2z_k+1)}+\frac{\tilde\delta(z_k)Q(z_k+1)}{2(z+z_k+1)(z_k+1)}\right]\prod_{i\neq k}\frac{1}{(z_k-z_i)(z_k+z_i+1)}\,,
 \end{split}
\end{equation} 
and
\begin{equation}\label{eq:nk}
\begin{split}
N_k&(z,\mathbf{z})=k_\mathcal{A}(z,z_k)\prod_{i\neq k}h_\mathcal{A}(z_k,z_i)\alpha(z_k)+k_\udt(z,z_k)\prod_{i\neq k}h_\udt(z_k,z_i)\tilde\delta(z_k)\,\\
 &=-\frac{2z+2}{2z+1}\left[\frac{\alpha(z_k)Q(z_k-1)}{(z+z_k+1)(2z_k+1)}+\frac{\tilde\delta(z_k)Q(z_k+1)}{2(z-z_k)(z_k+1)}\right]\prod_{i\neq k}\frac{1}{(z_k-z_i)(z_k+z_i+1)}\,.
 \end{split}
\end{equation} 
\section{FCR: Closed chain}\label{app:closed}
The relevant fundamental commutation relations that arise from the Yang-Baxter equation read
\begin{equation}
 [B(x),B(y)]=0\,,
\end{equation} 
\begin{equation}\label{eq:closedcom}
 A(x)B(y)=h(y,x)B(y)A(x)-k(y,x)B(x)A(y)\,,
\end{equation} 
with 
\begin{equation}
 h(x,y)=\frac{1+x-y}{x-y}\,,\qquad k(x,y)=\frac{1}{x-y}\,.
\end{equation}

\section{Wave function of the closed chain}\label{app:closedwave}
The off-shell Bethe vectors of the closed chain are given by, see e.g.~\cite{Essler2005},
\begin{equation}
 \prod_{i=1}^m B(z_i)|\Omega\rangle=\sum_{1\leq x_1<\ldots<x_m\leq L}\Phi(\mathbf{v},\mathbf{z},\mathbf{x})S_G\,e_{21}^{(x_1)}\cdots e_{21}^{(x_m)}S_G^{-1}|\Omega\rangle\,.
\end{equation} 
Here the wave function is of the form
\begin{equation}
 \Phi(\mathbf{v},\mathbf{z},\mathbf{x})=\sum_{\sigma\in S_m}{\rm A}^{\text{cl}}(z_{\sigma(1)},\ldots,z_{\sigma(m)})\prod_{i=1}^m\phi_{x_i}^{\text{cl}}(z_{\sigma(i)}, \mathbf{v})\,,
\end{equation} 
with the amplitude
\begin{equation}
 {\rm A}^{\text{cl}}(z_{\sigma(1)},\ldots,z_{\sigma(m)})=\prod_{1\leq k<l\leq m}\frac{z_{\sigma(k)}-z_{\sigma(l)}+1}{z_{\sigma(k)}-z_{\sigma(l)}}\,,
\end{equation} 
and the wave part
\begin{equation}
 \phi_{x}^{\text{cl}}(z_{\sigma(i)}, \mathbf{v})=\prod_{j=1}^{x-1}(z_{\sigma(i)}-v_j+1)\prod_{j=x+1}^L(z_{\sigma(i)}-v_j)\,.
\end{equation} 
\end{appendices}
\bibliographystyle{chetref}
\bibliography{openPBA}

\begin{thebibliography}{10}
\ifx\href\asklfhas\newcommand{\href}[2]{#2}\fi
\ifx\arxivref\asklfhas\newcommand{\arxivref}[2]{\href{http://arxiv.org/abs/#1}{#2}}\fi
\ifx\doiref\asklfhas\newcommand{\doiref}[2]{\href{http://dx.doi.org/#1}{#2}}\fi
\parskip 0pt
\normalsize

\bibitem{Baxter1987}
R.~J. Baxter,
\textit{``{P}erimeter {B}ethe ansatz''},
\doiref{10.1088/0305-4470/20/9/039}{J.~Phys.~A:~Math.~Gen. \textbf{20}, 2557
  (1987)\ignorespaces}\ignorespaces
\bibitem{Bethe1931}
H.~Bethe,
\textit{``{Z}ur {T}heorie der {M}etalle. {I}. {E}igenwerte und
  {E}igenfunktionen der linearen {A}tomkette''},
Zeitschrift~f{\"u}r~Ph \textbf{71}, 205 (1931)\ignorespaces\ignorespaces
\bibitem{Frassek2014}
R.~Frassek, N.~Kanning, Y.~Ko \& M.~Staudacher,
\textit{``{Bethe Ansatz for Yangian Invariants: Towards Super Yang-Mills
  Scattering Amplitudes}''},
\doiref{10.1016/j.nuclphysb.2014.03.015}{Nucl.~Phys. \textbf{B883}, 373
  (2014)\ignorespaces}\ignorespaces,
\normalsize{\texttt{\arxivref{1312.1693}{arXiv:1312.1693}}}\ignorespaces
\bibitem{Faddeev1996}
L.~Faddeev,
\textit{``{H}ow algebraic {B}ethe ansatz works for integrable model''},
\normalsize{\texttt{\arxivref{hep-th/9605187}{hep-th/9605187}}}\ignorespaces
\bibitem{Alcaraz1987}
F.~C. Alcaraz, M.~N. Barber, M.~T. Batchelor, R.~J. Baxter \& G.~R.~W. Quispel,
\textit{``{S}urface exponents of the quantum {XXZ}, {A}shkin-{T}eller and
  {P}otts models''},
\doiref{10.1088/0305-4470/20/18/038}{J.~Phys.~A:~Math.~Gen. \textbf{20}, 6397
  (1987)\ignorespaces}\ignorespaces
\bibitem{Gaudin2014}
M.~{Gaudin},
\textit{``{The Bethe Wavefunction}''},
Cambridge University Press (2014)\ignorespaces
\bibitem{Sklyanin:1988yz}
E.~K. Sklyanin,
\textit{``{Boundary Conditions for Integrable Quantum Systems}''},
\doiref{10.1088/0305-4470/21/10/015}{J.~Phys. \textbf{A21}, 2375
  (1988)\ignorespaces}\ignorespaces
\bibitem{Cherednik1984a}
I.~V. Cherednik,
\textit{``{Factorizing Particles on a Half Line and Root Systems}''},
\doiref{10.1007/BF01038545}{Theor.~Math.~Phys. \textbf{61}, 977
  (1984)\ignorespaces}\ignorespaces
\bibitem{Olshanskii1992}
G.~Olshanskii,
\textit{``{T}wisted {Y}angians and infinite-dimensional classical {L}ie
  algebras''},
in \textit{``Quantum groups''},
Springer (1992)\ignorespaces,
104--119\ignorespaces
\bibitem{Molev:2007}
A.~Molev,
\textit{``{Yangians and Classical Lie Algebras}''},
American Mathematical Society (2007)\ignorespaces
\bibitem{MacKay2005}
N.~J. MacKay,
\textit{``{Introduction to Yangian symmetry in integrable field theory}''},
\doiref{10.1142/S0217751X05022317}{Int.~J.~Mod.~Phys. \textbf{A20}, 7189
  (2005)\ignorespaces}\ignorespaces,
\normalsize{\texttt{\arxivref{hep-th/0409183}{hep-th/0409183}}}\ignorespaces,
in \textit{``{ESI Workshop on String Theory on Non-Compact and Time-Dependent
  Backgrounds Vienna, Austria, June 7-18, 2004}''},
7189-7218\ignorespaces
\bibitem{Zamolodchikov1990}
A.~B. Zamolodchikov,
\textit{``{Factorized S matrices and lattice statistical systems}''},
Soviet~Scientific~Reviews \textbf{2}, 1 (1990)\ignorespaces\ignorespaces
\bibitem{Wang2002}
Y.-S. Wang,
\textit{``{T}he scalar products and the norm of {B}ethe eigenstates for the
  boundary {XXX} {H}eisenberg spin-1/2 finite chain''},
\doiref{http://dx.doi.org/10.1016/S0550-3213(01)00610-1}{Nuclear~Physics~B
  \textbf{622}, 633  (2002)\ignorespaces}\ignorespaces
\bibitem{Kitanine:2007bi}
N.~Kitanine, K.~K. Kozlowski, J.~M. Maillet, G.~Niccoli, N.~A. Slavnov \&
  V.~Terras,
\textit{``{Correlation functions of the open XXZ chain I}''},
\doiref{10.1088/1742-5468/2007/10/P10009}{J.~Stat.~Mech. \textbf{0710}, P10009
  (2007)\ignorespaces}\ignorespaces,
\normalsize{\texttt{\arxivref{0707.1995}{arXiv:0707.1995}}}\ignorespaces
\bibitem{Essler2005}
F.~H. Essler, H.~Frahm, F.~G{\"o}hmann, A.~Kl{\"u}mper \& V.~E. Korepin,
\textit{``{T}he one-dimensional {H}ubbard model''},
Cambridge University Press (2005)\ignorespaces
\bibitem{Chicherin2014b}
D.~Chicherin, S.~Derkachov \& R.~Kirschner,
\textit{``{Yang-Baxter operators and scattering amplitudes in N=4
  super-Yang-Mills theory}''},
\doiref{10.1016/j.nuclphysb.2014.02.016}{Nucl.~Phys. \textbf{B881}, 467
  (2014)\ignorespaces}\ignorespaces,
\normalsize{\texttt{\arxivref{1309.5748}{arXiv:1309.5748}}}\ignorespaces
\bibitem{Kanning:2014cca}
N.~Kanning, Y.~Ko \& M.~Staudacher,
\textit{``{Graßmannian integrals as matrix models for non-compact Yangian
  invariants}''},
\doiref{10.1016/j.nuclphysb.2015.03.011}{Nucl.~Phys. \textbf{B894}, 407
  (2015)\ignorespaces}\ignorespaces,
\normalsize{\texttt{\arxivref{1412.8476}{arXiv:1412.8476}}}\ignorespaces
\bibitem{Arkani-Hamed2012}
N.~Arkani-Hamed, J.~L. Bourjaily, F.~Cachazo, A.~B. Goncharov, A.~Postnikov \&
  J.~Trnka,
\textit{``{Scattering Amplitudes and the Positive Grassmannian}''},
Cambridge University Press (2012)\ignorespaces
\bibitem{offdiag}
Y.~Wang, W.-L. Yang, J.~Cao \& K.~Shi,
\textit{``{Off-Diagonal Bethe Ansatz for Exactly Solvable Models}''},
Springer (2015)\ignorespaces
\bibitem{Tsuchiya1998a}
O.~Tsuchiya,
\textit{``{D}eterminant formula for the six-vertex model with reflecting
  end''},
\doiref{10.1063/1.532606}{J.~Math.~Phys. \textbf{39}, 5946
  (1998)\ignorespaces}\ignorespaces,
\normalsize{\texttt{\arxivref{solv-int/9804010}{solv-int/9804010}}}\ignorespaces
\bibitem{Gier2011}
J.~de~Gier, W.~Galleas \& M.~Sorrell,
\textit{``{Multiple integral formula for the off-shell six vertex scalar
  product}''},
\normalsize{\texttt{\arxivref{1111.3712}{arXiv:1111.3712}}}\ignorespaces
\bibitem{Galleas2014}
W.~Galleas \& J.~Lamers,
\textit{``{Reflection algebra and functional equations}''},
\doiref{10.1016/j.nuclphysb.2014.07.016}{Nucl.~Phys. \textbf{B886}, 1003
  (2014)\ignorespaces}\ignorespaces,
\normalsize{\texttt{\arxivref{1405.4281}{arXiv:1405.4281}}}\ignorespaces
\bibitem{Dorey1999}
P.~Dorey \& R.~Tateo,
\textit{``{Anharmonic oscillators, the thermodynamic Bethe ansatz, and
  nonlinear integral equations}''},
\doiref{10.1088/0305-4470/32/38/102}{J.~Phys. \textbf{A32}, L419
  (1999)\ignorespaces}\ignorespaces,
\normalsize{\texttt{\arxivref{hep-th/9812211}{hep-th/9812211}}}\ignorespaces
\bibitem{Bazhanov2003}
V.~V. Bazhanov, S.~L. Lukyanov \& A.~B. Zamolodchikov,
\textit{``{Higher level eigenvalues of Q operators and Schroedinger
  equation}''},
\doiref{10.4310/ATMP.2003.v7.n4.a4}{Adv.~Theor.~Math.~Phys. \textbf{7}, 711
  (2003)\ignorespaces}\ignorespaces,
\normalsize{\texttt{\arxivref{hep-th/0307108}{hep-th/0307108}}}\ignorespaces
\bibitem{Frassek:2015mra}
R.~Frassek \& I.~M. Szecsenyi,
\textit{``{Q-operators for the open Heisenberg spin chain}''},
\doiref{10.1016/j.nuclphysb.2015.10.010}{Nucl.~Phys. \textbf{B901}, 229
  (2015)\ignorespaces}\ignorespaces,
\normalsize{\texttt{\arxivref{1509.04867}{arXiv:1509.04867}}}\ignorespaces
\bibitem{Arkani-Hamed2015}
N.~Arkani-Hamed, J.~L. Bourjaily, F.~Cachazo, A.~Postnikov \& J.~Trnka,
\textit{``{On-Shell Structures of MHV Amplitudes Beyond the Planar Limit}''},
\doiref{10.1007/JHEP06(2015)179}{JHEP \textbf{1506}, 179
  (2015)\ignorespaces}\ignorespaces,
\normalsize{\texttt{\arxivref{1412.8475}{arXiv:1412.8475}}}\ignorespaces
\bibitem{Drummond:2010uq}
J.~M. Drummond \& L.~Ferro,
\textit{``{The Yangian origin of the Grassmannian integral}''},
\doiref{10.1007/JHEP12(2010)010}{JHEP \textbf{1012}, 010
  (2010)\ignorespaces}\ignorespaces,
\normalsize{\texttt{\arxivref{1002.4622}{arXiv:1002.4622}}}\ignorespaces
\bibitem{Frassek2016}
R.~Frassek \& D.~Meidinger,
\textit{``{Yangian-type symmetries of non-planar leading singularities}''},
\doiref{10.1007/JHEP05(2016)110}{JHEP \textbf{1605}, 110
  (2016)\ignorespaces}\ignorespaces,
\normalsize{\texttt{\arxivref{1603.00088}{arXiv:1603.00088}}}\ignorespaces
\bibitem{Berenstein:2005vf}
D.~Berenstein \& S.~E. Vazquez,
\textit{``{Integrable open spin chains from giant gravitons}''},
\doiref{10.1088/1126-6708/2005/06/059}{JHEP \textbf{0506}, 059
  (2005)\ignorespaces}\ignorespaces,
\normalsize{\texttt{\arxivref{hep-th/0501078}{hep-th/0501078}}}\ignorespaces
\bibitem{Drukker:2012de}
N.~Drukker,
\textit{``{Integrable Wilson loops}''},
\doiref{10.1007/JHEP10(2013)135}{JHEP \textbf{1310}, 135
  (2013)\ignorespaces}\ignorespaces,
\normalsize{\texttt{\arxivref{1203.1617}{arXiv:1203.1617}}}\ignorespaces
\bibitem{deLeeuw:2015hxa}
M.~de~Leeuw, C.~Kristjansen \& K.~Zarembo,
\textit{``{One-point Functions in Defect CFT and Integrability}''},
\doiref{10.1007/JHEP08(2015)098}{JHEP \textbf{1508}, 098
  (2015)\ignorespaces}\ignorespaces,
\normalsize{\texttt{\arxivref{1506.06958}{arXiv:1506.06958}}}\ignorespaces
\end{thebibliography}

\noindent\rule{6cm}{0.4pt}

\texttt{Contact: frassek@ihes.fr}

\end{document}